\documentclass[apjl,iop]{emulateapj}
\slugcomment{To be published in the Astrophysical Journal Letters} 
\usepackage[table]{xcolor}
\usepackage[normalem]{ulem}
\usepackage{rotating}

\shortauthors{Pasham \& Strohmayer}
\begin{document}
\title{Can the 62 day X-ray period of the Ultraluminous X-ray source M82 X-1 \\ be due to a precessing accretion disk?}
\author{Dheeraj R. Pasham\altaffilmark{1,2}, Tod E. Storhmayer\altaffilmark{2}}
\affil{$^{1}$Astronomy Department, University of Maryland, College Park, MD 20742; email: dheeraj@astro.umd.edu; richard@astro.umd.edu \\ 
$^{2}$Astrophysics Science Division, NASA's GSFC, Greenbelt, MD 20771; email: tod.strohmayer@nasa.gov \\
{\it Received 2013 June 23; Accepted 2013 August 1}
}

\begin{abstract}
We have analyzed all archival {\it RXTE/PCA} monitoring observations
of the ultraluminous X-ray source (ULX) M82 X-1 in order to study the
properties of its 62 day X-ray period (Kaaret \& Feng 2007). Based on
its high coherence it has been argued that the observed period is the
orbital period of the binary. Utilizing a much longer data set than in
previous studies we find: (1) The phase-resolved X-ray (3-15 keV)
spectra -- modeled with a thermal accretion disk and a power-law --
suggest that the accretion disk's contribution to the total flux is
strongly modulated with phase. (2) Suggestive evidence for a sudden
phase shift--of approximately 0.4 in phase (25 days)--between the
first and the second halves of the light curve separated by roughly
1000 days. If confirmed, the implied timescale to change the period is
$\sim$ 10 yrs, which is exceptionally fast for an orbital phenomenon.
These two independent pieces of evidence are consistent with the
periodicity being due to a precessing accretion disk, similar to the
super-orbital periods observed in systems like Her X-1, LMC X-4, and
SS433. However, the timing evidence for a change in the period needs
to be confirmed with additional observations. This should be possible
with further monitoring of M82 with instruments such as the {\it
Swift} X-ray telescope (XRT).
\end{abstract}


\vfill\eject
\pagebreak
\newpage

\section{Introduction}
Ultraluminous X-ray sources (ULXs) are bright, point-like X-ray
sources in nearby galaxies with apparent luminosities in the range of
a few$\times$10$^{39-41}$ ergs s$^{-1}$ (Fabbiano 1989; Swartz et al. 2011). They are mysterious in the sense that
their energy output exceeds the Eddington limit for stellar-mass black
holes (mass range of 3-20 M$_{\odot}$) (see review by Miller \&
Colbert 2004). They could be stellar-mass black
holes undergoing super-Eddington accretion and/or emission (King et
al. 2001; Begelman 2002; K{\"o}rding et al. 2002; Gladstone et
al. 2009) or the missing class of intermediate-mass black holes (mass
range: few$\times$(100-1000) M$_{\odot}$) accreting at sub-Eddington
rates (Colbert \& Mushotzky 1999).

With a maximum X-ray luminosity of approximately 10$^{41}$ ergs
s$^{-1}$ (Kaaret et al. 2009) M82 X-1 is a remarkably bright ULX. Its
average X-ray luminosity of roughly 5$\times$10$^{40}$ ergs s$^{-1}$
combined with quasi-periodic oscillations (QPOs) in the frequency
range of 0.04-0.2 Hz suggests that it may contain an intermediate-mass
black hole of roughly 100-1000 M$_{\odot}$ (Strohmayer \& Mushotzky
2003; Hopman et al. 2004; Portegies Zwart et al. 2005; Dewangan et
al. 2006; Mucciarelli et al. 2006; Pasham \& Strohmayer 2013). Another
intriguing property is that its X-ray intensity varies regularly with
a period of 62 days (Kaaret et al. 2006; Kaaret \& Feng 2007, KF07
hereafter). Such long, periodic X-ray modulations have been seen from
two other ULXs, NGC 5408 X-1 (115 days and 243 days: Strohmayer 2009;
Han et al. 2012; Pasham \& Strohmayer 2013) and HLX ESO 243-39 (375
days: e.g., Servillat et al. 2011). The 62 day period of M82 X-1 has
been claimed to be the orbital period of the black hole binary system
(KF07). Here, we study the properties of this period using new data
and present evidence that this modulation may instead be due to the
precessing accretion disk of the black hole.

\section{Data Primer}
The data used herein was obtained with the Rossi X-ray Timing
Explorer's ({\it RXTE}'s) proportional counter array ({\it PCA})
operating in the {\it GoodXenon} data acquisition mode. We used data
from the monitoring program beginning 2004 September 2 until 2009
December 30 (1945 days), during which M82 was observed roughly once
every three days (2-3 ks per observation).

We used data from all active proportional counter units (PCUs). For
faint sources (net count rates $\lessapprox$ 20 counts sec$^{-1}$)
like M82, the {\it PCA} data analysis guide provided by {\it RXTE}'s
Guest Observer Facility
(GOF, http://heasarc.nasa.gov/docs/xte/recipes/layers.html) suggests using
only the top Xenon layer to maximize the signal to noise
ratio. Therefore, we screened our data to include only events from the
top layer (layer-1) with both anode chains (Left and Right). In
addition, we imposed the following standard filter on the data: ELV
$>$ 10.0 \&\& OFFSET $<$ 0.02 \&\& (TIME\_SINCE\_SAA $<$ 0 $||$
TIME\_SINCE\_SAA $>$ 30) \&\& ELECTRON2 $<$ 0.1. Finally, we used the
latest SAA history and background model files for our analysis. 


%
%
\begin{figure*}
  \begin{center}
 \includegraphics[width=18cm]{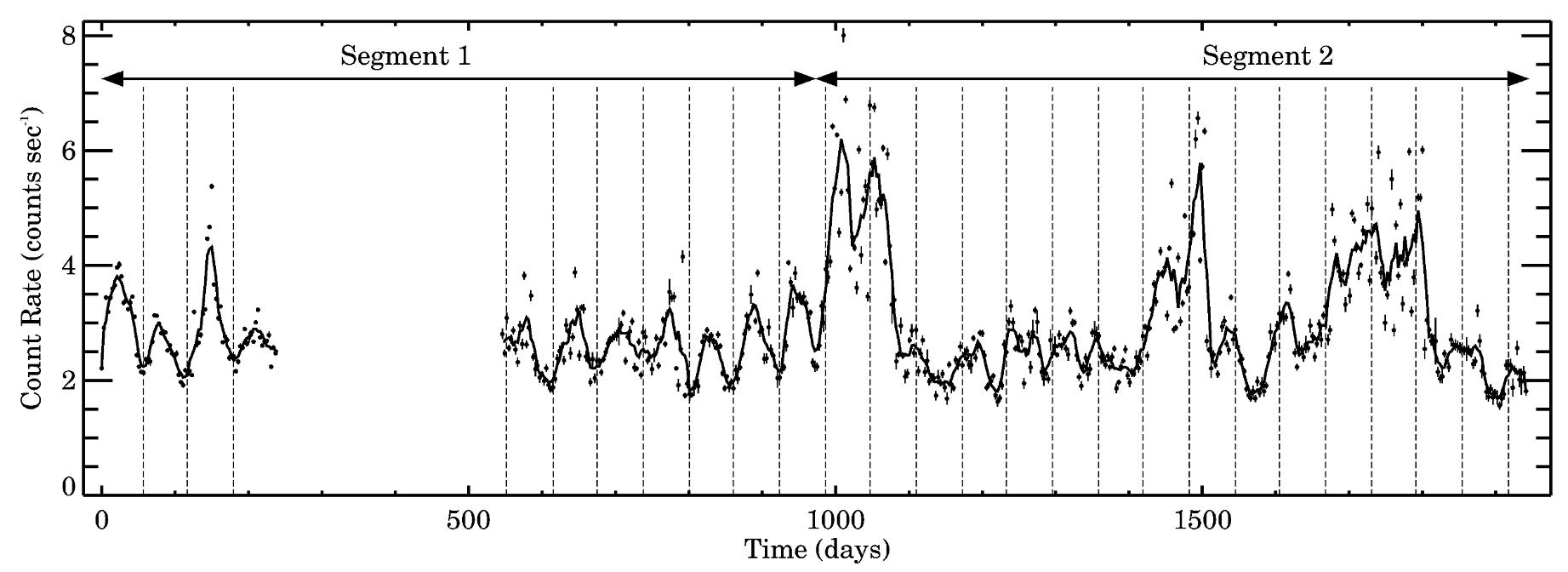}    
       \caption[Figure1]{Complete {\it RXTE/PCA} binned X-ray (3-15 keV)
light curve of M82 ({\it solid points}) along with the running average
({\it solid curve}) that traces the overall X-ray variability of
M82. The error bars on the individual data points are also shown. The
start time of the light curve is 2004 September 2, 14:26:14.757
UTC. The bin size is 3 days. The vertical lines show the expected
minima of the X-ray modulation assuming the 62 day period remains
constant throughout the data. The two segments represent data
before and after the first major flare that occurs around day 1000. }
     \label{fig:fig1}
  \end{center}
\end{figure*}


The PCA observations were divided amongst six proposals ({\it RXTE}
proposal IDs: P20303, P90121, P90171, P92098, P93123, P94123). We used
the {\it rex} script provided by {\it RXTE}'s GOF to extract
individual light curves and energy spectra of the source as well as
the background. In addition to the filters described above we only
used data from channels 0-35 which translates to X-ray events in the
energy range 3-15 keV.

\section{Results}
\subsection{Timing Analysis}
From each individual observation we extracted an average\footnote{The
mean was taken over all active PCUs using PCU normalizations given by
the {\it RXTE} data analysis guide:
http://heasarc.nasa.gov/docs/xte/recipes/pcu\_combine.html.},
background-subtracted, count rate. This resulted in a total of 810
measurements distributed over 1945 days. The complete {\it RXTE/PCA}
3-15 keV light curve of M82 is shown in Figure 1. While the earlier
work by KF07 used only the data from day 0 until roughly day 900,
i.e., essentially segment 1 of Figure 1, this work includes the entire
{\it RXTE/PCA} monitoring data of M82.

As an initial test for the stability of the period, we over-plotted
vertical lines uniformly separated by 62 days\footnote{The best-fit
period reported by KF07 was 62$\pm$0.3 days.} and coincident with the
expected minima of the light curve assuming this period is constant
(dashed vertical lines in Figure 1). It is clear even by eye that
while the vertical lines are coincident with the light curve's minima
until the large flare occurring around day 1000, they are offset
thereafter. The location of the minima were estimated as follows. We
folded the first four cycles of the data at a period of 62 days, i.e.,
data from day 0 - day 240. We then fit this folded light curve with a
model that includes two Fourier components (the fundamental and first
harmonic), i.e., I = A + Bsin2$\pi$($\phi$-$\phi_{0}$) +
Csin4$\pi$($\phi$-$\phi_{1}$). The folded light curve ({\it solid
points}) along with the best-fit function ({\it solid curve}) is shown
in the left panel of Figure 2. The best-fit model parameters are A =
2.74$\pm$0.01, B = 0.38$\pm$0.01, $\phi_{0}$ =1.14$\pm$0.01, C =
0.12$\pm$0.01, $\phi_{1}$ = 1.04$\pm$0.01 while the best-fit
$\chi^{2}$ value was 1.4 for 1 degree of freedom. If the 62 day
modulation were constant throughout the monitoring program then the
minima of the best-fit model should track the light curve's minima.


%
%
\begin{figure*}
  \begin{center}
 \includegraphics[width=18cm]{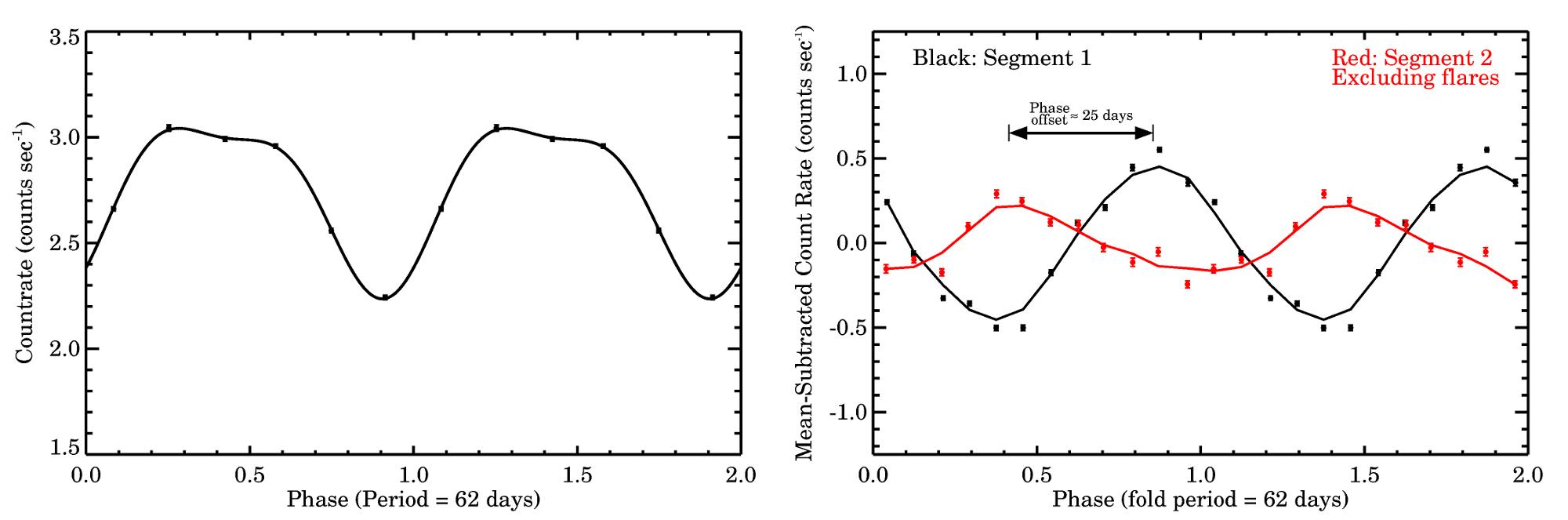}    
       \caption[Figure2]{Left Panel: Folded X-ray (3-15 keV) light curve of
M82 ({\it solid points}) along with the best-fit sinusoid curve ({\it
solid curve}) using only data from day 0 until day 240 (see text). 6
bins per cycle were used and two cycles are shown to guide the
eye. Right Panel: Mean-subtracted, folded X-ray (3-15 keV) light curves
of M82 during segment 1 (black) and segment 2 excluding the flares
(red). In each case a total of 12 bins per cycle were used and two
cycles are shown for clarity. The error bars on the individual phase
bins are also shown. The solid curves represent the running average
over three neighboring bins. A phase offset of $\approx$ 25 days
between the two portions of the light curve is evident.}
     \label{fig:fig2}
  \end{center}
\end{figure*}


Another way to assess this phase change is to separately fold the
segments of the light curve before and after the first large flare
(segment 1 and 2 as indicated in Figure 1), at the period of 62
days. Therefore, we divided the complete light curve into two
segments: 1) prior to the large flare and 2) after the large flare
excluding the data during flares. For the first segment we used data
from day 0 to day 976 where day 976 represents roughly the epoch of
the onset of the flare (see Figure 1). For the second segment we used
all the data from day 976 until the end of the light curve except for
the flares. We then transformed the two segments of the light curve to
have the same start time. This is essential as a phase difference
between the start times of the two segments can manifest as an offset
between their folded light curves. We then folded the two segments
separately at a period of 62 days\footnote{Note that we have
constructed a Lomb-Scargle periodogram (Scargle 1982; Horne \&
Baliunas 1986) of segment 2 and find evidence -- although weaker
compared to segment 1 -- for a power spectral peak that is consistent
with a period of 62 days.} as found in the earlier work by KF07.  The
two folded light curves (offset to have zero mean) are shown in the
right panel of Figure 2. Clearly there is a significant phase offset
of roughly 0.4 cycles -- equivalent to 0.4$\times$62 days $\approx$ 25
days -- between the two portions of the light curve.

It is possible that this phase difference is due to an incorrect
choice of the fold period. Considering the uncertainty in the period
reported by KF07, the actual value of the period can be in the range
($90\%$ confidence) $62 \pm 0.3$ days. Therefore, we repeated the
analysis using various fold periods between 61.7 and 62.3 days. We
find that the lag is significant in all the cases with the lag varying
from roughly 20 to 30 days (see Figure 3). However, if we relax the
confidence interval on the best-fit period, we find that one can
obtain essentially zero lag between the two segments with a fold
period of 60.6 days. We note that this value is 4.7 times the quoted
uncertainty, that is, $(1.4/0.3)=4.7$, away from the best-fit period
of 62 days. Since a $90\%$ confidence region is $\approx 1.6\sigma$
(assuming gaussian statistics) from the best value, then one has to go
$1.6\times4.7 = 7.5\sigma$ from the best period (62 days) in order to
cancel the inferred lag.  This supports the presence of a real phase
shift, but due to the relatively modest number of overall cycles
present in the data, a confirmation of a varying period would still be
important.


%
%
\begin{figure*}
  \begin{center}
 \includegraphics[width=18cm]{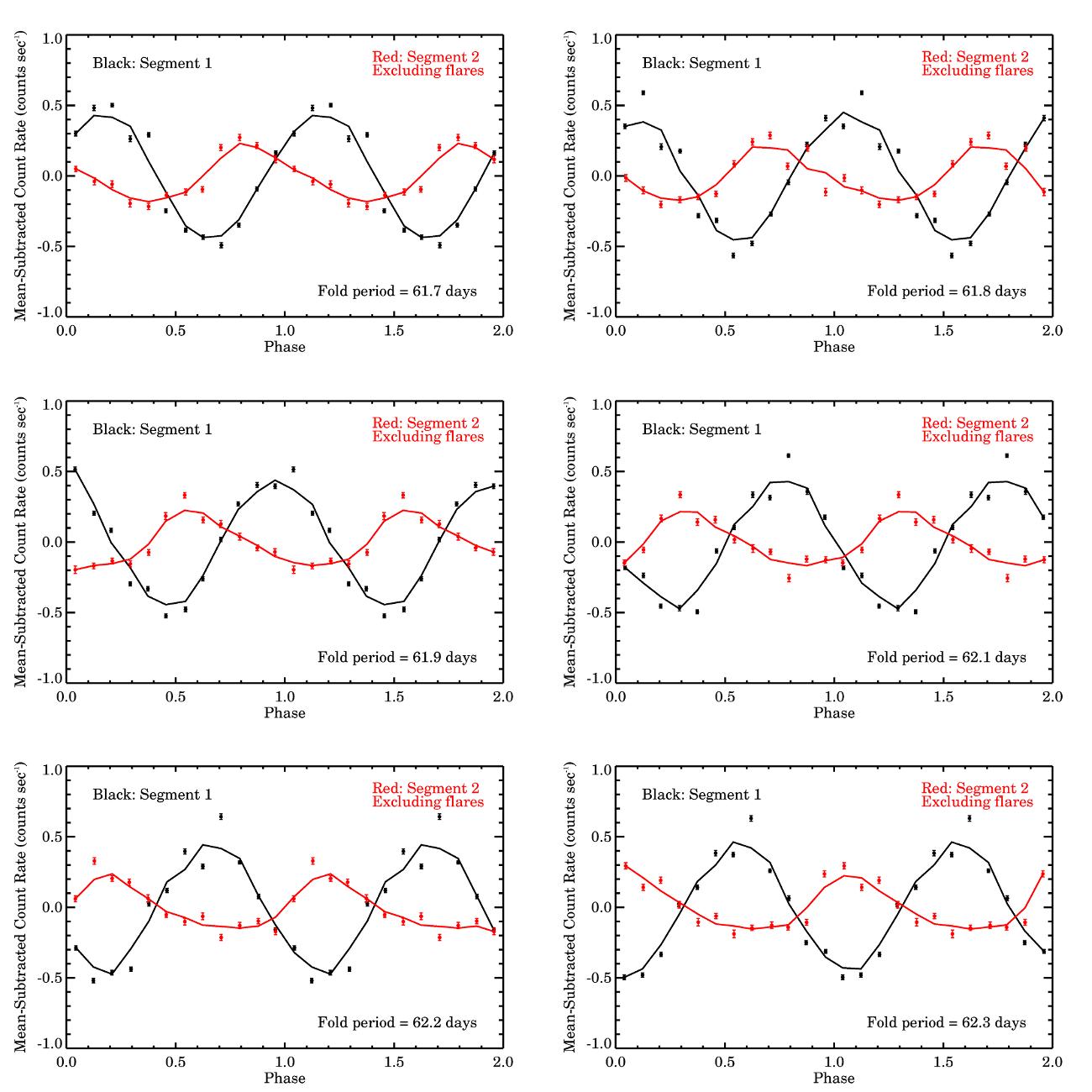}    
       \caption[Figure3]{Same as the right panel of Figure 2 but using various fold periods within the error bar reported by Kaaret \& Feng (2007). The fold period used is indicated in each panel (bottom-right). A significant ``phase-offset'' is evident in each case.}
     \label{fig:fig3}
  \end{center}
\end{figure*}


\section{Energy Spectral Analysis}
For the purposes of extracting phase-resolved energy spectra we used
only data from day zero until prior to the first large flare around
day 1000, i.e., segment 1 of Figure 1. We made this choice for the
following reasons: (1) the 62 day modulation is highly coherent during
this portion of the data and (2) the flaring in the second segment
likely introduces additional state-related spectral variations (Feng
\& Kaaret 2010) which could mask any purely phase-related changes.

We then extracted the energy spectra of the source and background from
each observation in segment 1. Using the tool {\it pcarsp} we created
responses separately for each observation. If multiple PCUs were
active in any given observation, we first obtained the source, the
background spectra and the responses from individual PCUs and then
combined them to have a single source spectrum, a background spectrum
and a response (the individual PCU responses were weighted according
to the background-subtracted source counts) per observation. We then
divided these observations into six equal-sized phase bins of size 1/6
using a period of 62 days. Using the FTOOL {\it sumpha} we combined
all the source and the background energy spectra in a given phase bin
to obtain six average phase-resolved source and background energy
spectra. Similarly using the FTOOLs {\it addrmf} and {\it addarf}, we
created six weight-averaged response matrices and the ancillary
response functions, respectively. For each of these twelve response
files (six RMFs and six ARFs) weights were assigned according to the
total number of background-subtracted counts in a given
observation. We then binned the energy spectra to ensure a minimum of
30 counts in each spectral bin.



\begin{figure}[ht]
\includegraphics[width=3.5in,height=4.25in]{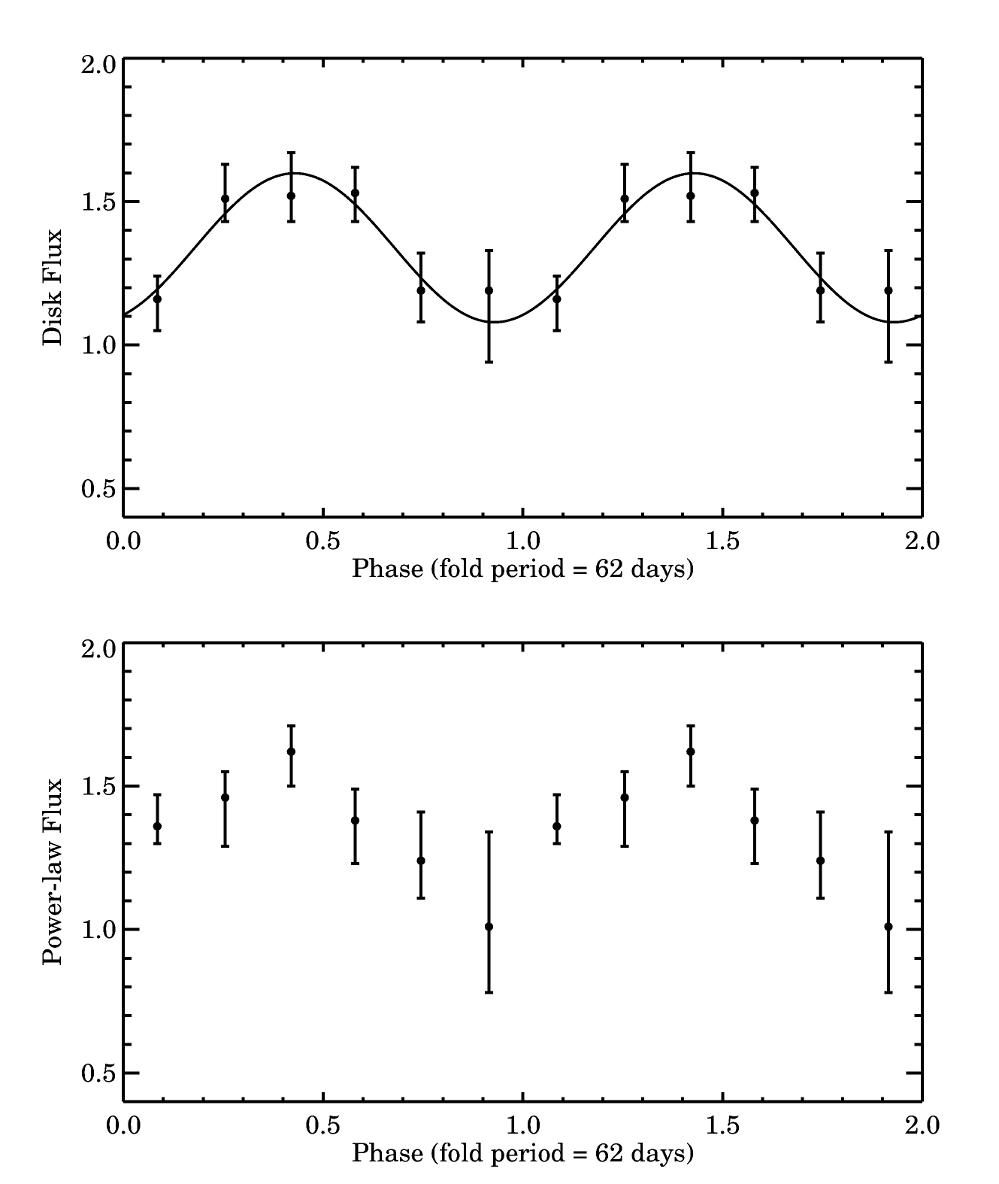}    

\caption{The phase-resolved X-ray (3-15 keV) disk flux
({\it y-axis} in the top panel) and the power-law flux ({\it
y-axis} in the bottom panel) as a function of the 62 day phase
({\it x-axis}). The flux units are 10$^{-11}$ ergs s$^{-1}$
cm$^{-2}$. In each case two cycles are shown for clarity. In order to
guide the eye, the best-fitting sinusoid curves ({\it solid}) defined
as A + B*Sin[2$\pi$(phase-constant$_{0}$)] are also indicated in the
top panel. The error bars represent 1$\sigma$ uncertainty on the
flux. Clearly, the disk component is strongly modulated. There is only
weak evidence for power-law modulation.}
\label{fig:figure4}
\end{figure}



We modeled each energy spectrum with a blackbody disk, a power-law
model, and a gaussian component to account for the weakly broadened Fe
K$\alpha$ line. We used the {\it XSPEC} (Arnaud 1996) spectral fitting
package to fit the spectra. In terms of {\it XSPEC} models, we
used {\it phabs*(diskpn + gauss + pow)}. The spectral resolution of
the data does not allow us to constrain the Gaussian parameters but it
is required for a good fit. Therefore, we fixed the centroid energy
and the width of the iron line at 6.55 keV and 0.33 keV,
respectively. We obtained these values from earlier work using
high-resolution {\it Suzaku} and {\it XMM-Newton} observations of M82
X-1 (Strohmayer \& Mushotzky 2003; Caballero-Garc{\'{\i}}a 2011). We
find that in the energy range 3-15 keV this model fits the data
well, giving reduced $\chi^{2}$ values in the range of 0.5-1.1 for 23
degrees of freedom. All the best-fit model parameters are indicated in
Table 1. Figure 4 shows the value of the disk and the power-law fluxes
as a function of the phase. Clearly, the disk flux varies with phase.


\begin{table*}
  \begin{flushleft}
\centering
  \caption{Summary of the phase-resolved energy spectral modeling of M82. Best-fitting parameters using the {\it phabs$\ast$(diskpn+gauss+pow)} model are shown. }\label{Table1} 
\vspace{0.05cm}
{
    \begin{tabular}[t]{lccccccc}
\\
    \hline\hline \\	
& & & {\it phabs$\ast$(diskpn+gauss+pow)}: & & & \\ 
\\
    \hline\hline \\

   Phase$^{a}$ 			& 0.085			& 0.255			& 0.420			 & 0.58			  & 0.745		   & 0.915 		    \\
	\\
    \hline \\
T$^{b}_{max}$   		& 2.2$^{+0.2}_{-0.2}$   &  2.1$^{+0.2}_{-0.1}$  &   2.1$^{+0.2}_{-0.1}$  &   2.2$^{+0.2}_{-0.2}$  &   2.4$^{+0.3}_{-0.2}$  &   2.8$^{+0.2}_{-0.4}$  \\
\\
N$^{c}_{disk}$			& 5.7$^{+2.7}_{-1.8}$  & 9.1$^{+3.6}_{-2.7}$    &  10.0$^{+3.9}_{-2.9}$  & 7.5$^{+3.0}_{-2.2}$    &  4.1$^{+2.0}_{-1.2}$   & 2.0$^{+1.0}_{-0.5}$    \\ 
\\
N$^{d}_{gauss}$			& 3.9$^{+0.7}_{-0.7}$   & 4.1$^{+0.8}_{-0.8}$   &  4.4$^{+0.8}_{-0.8}$   & 4.2$^{+0.8}_{-0.8}$    &  3.9$^{+0.7}_{-0.7}$   & 4.3$^{+0.8}_{-0.8}$     \\ 
\\
$\Gamma$$^{e}$			&  2.0$^{+0.2}_{-0.2}$	& 1.8$^{+0.2}_{-0.2}$  & 1.8$^{+0.2}_{-0.2}$  & 1.8$^{+0.2}_{-0.2}$ & 2.1$^{+0.2}_{-0.2}$  & 2.5$^{+0.6}_{-0.3}$	     \\
\\
N$^{e}_{powlaw}$		&  5.4$^{+1.6}_{-1.5}$  & 3.8$^{+1.6}_{-1.6}$  &  4.6$^{+1.7}_{-1.7}$   & 4.1$^{+1.6}_{-1.5}$   &  6.3$^{+1.9}_{-1.6}$  & 9.6$^{+6.6}_{-2.6}$    \\   
\\
F$^{f}_{X}$			&2.60$^{+0.02}_{-0.07}$ &3.10$^{+0.02}_{-0.06}$ &3.22$^{+0.05}_{-0.11}$  & 2.98$^{+0.03}_{-0.11}$ & 2.48$^{+0.02}_{-0.13}$ & 2.26$^{+0.06}_{-0.05}$      \\
\\  
F$^{f}_{Disk}$			& 1.16$^{+0.08}_{-0.11}$&1.51$^{+0.12}_{-0.08}$ &1.52$^{+0.15}_{-0.09}$  & 1.53$^{+0.09}_{-0.10}$ & 1.19$^{+0.13}_{-0.11}$ & 1.19$^{+0.14}_{-0.25}$   \\
\\
F$^{f}_{Power}$			& 1.36$^{+0.11}_{-0.06}$&1.46$^{+0.09}_{-0.17}$ &1.62$^{+0.09}_{-0.12}$  & 1.38$^{+0.11}_{-0.15}$ & 1.24$^{+0.17}_{-0.13}$ & 1.01$^{+0.33}_{-0.23}$    \\
\\
$\chi^2$/dof			& 14/23		& 11/23		& 18/23		 & 24/23		  & 21/23		   & 18/23			 \\
\\
 \hline\hline
    \end{tabular}
\\
}
  \end{flushleft}

{\scriptsize
$^{a}${We obtained six phase-resolved energy spectra where each spectrum is an average of all data within 1/6$^{th}$ of the phase bin.}
$^{b}${Accretion disk temperature in keV. The inner disk radius was fixed at 6GM/c$^{2}$.}
$^{c}${Normalization ($\times$10$^{-7}$) of the {\it diskpn} component.}
$^{d}${Normalization ($\times$10$^{-5}$) of the {\it gaussian} component (Fe K$\alpha$ emission line).}
$^{e}${Index ($\Gamma$) and the normalization ($\times$10$^{-3}$) of the power-law component of the energy spectrum.}
$^{f}${Total X-ray flux (F$_{X}$), Disk flux (F$_{Disk}$) and the power-law flux (F$_{Power}$) of the energy spectrum in the energy range of 3-15 keV (units are 10$^{-11}$ ergs s$^{-1}$ cm$^{-2}$).}
The column density of hydrogen along the line of sight was fixed at 1.1$\times$10$^{22}$cm$^{-2}$ -- the best-fitting value found by Feng \& Kaaret (2010) using the high spatial resolution {\it Chandra} data.
}
\end{table*}


\section{Discussion}
The phase offset noted above ($\approx$ 0.4 cycles or about 25 days)
occurs over roughly 1000 days. This corresponds to a characteristic
timescale of 1/(0.4/1000) or $\sim$ 10 yrs. This is unusually fast for
an orbital phenomenon. The typical values of evolution timescales of
orbits of accreting compact binaries (neutron star or black hole
binaries) are a few$\times$10$^{6}$ yrs (e.g., Verbunt 1993; Levine et
al. 2000; Wolff et al. 2009; Jain et al. 2010 and
references therein). This suggests that the evolutionary timescale of
the phenomenon associated with the 62 day period may be $\sim$
10$^{5}$ times faster than typical. Periods longer than the orbital
period have been detected from numerous compact binaries (e.g., Kotze
\& Charles 2012, KC12 hereafter; Wen et al. 2006, W06
hereafter). These are known as super-orbital periods and are ascribed
to disk precession (e.g., Katz 1973; Pringle 1996; Ogilvie \& Dubus
2001) in addition to several other mechanisms (e.g., KC12). A
characteristic feature of super-orbital periods is that they are often
accompanied by sudden changes in coherence, either in the period or
the phase, similar to the suggested behavior reported here from M82
(e.g., Clarkson et al. 2003; KC12).

On the other hand, numerous systems exhibit relatively stable
super-orbital periods. These include Her X-1 with a period between
33-37 days (e.g., Leahy \& Igna 2010; Figure 16 of KC12), LMC X-4 with
$\sim$ 30 days (W06; Figure 5 of KC12), and SS433 with a period of
$\sim$ 162 days (W06; Figure 7 of KC12).  2S 0114+650 also shows a
stable super-orbital period (Figure 8 of KC12) but this may not be due
to a precessing accretion disk (e.g., Farrell et al. 2006). At least
in Her X-1 phase shifts are known to occur (see, for example, Figure 9
of Clarkson et al. 2003) just before the onset of the so-called
anomalous low state. 

Moreover, if this modulation is indeed due to a precessing accretion
disk one expects the X-ray flux originating from the disk to vary with
a period of 62 days. This is simply due to the fact that as the
accretion disk precesses its projected area on the sky varies with the
phase of the precession period. The observed dependence of the disk
flux with the phase of the 62 day period is consistent with this idea
(see Figure 4).

One of the controversies surrounding M82 X-1 is whether it hosts an
intermediate-mass or stellar-mass black hole. If the 62 day period is
indeed due to relatively stable precession of the accretion disk --
perhaps due to radiation induced warping -- then probing the warp
structure can, in principle, give us some insight into the mass
question. Given the accretion efficiency of the black hole
($\epsilon$) and the ratio of the viscosity in the normal to the
planar direction ($\eta$), Pringle (1996) derived the radius $R$
beyond which the disk warps,

\begin{equation}
\frac{R}{R_{s}} \geq \left ( \frac{2\sqrt{2}\pi\eta}{\epsilon} \right)^{2}
\end{equation}

where $R_{s}$ is the Schwarzschild radius ($2GM/c^{2}$). The value of
$\eta$ is $\sim$ 1 (Pringle 1996). As noted earlier, M82 X-1 has an
average luminosity of 5$\times$10$^{40}$ ergs s$^{-1}$. Assuming
isotropic emission the relation connecting the mass of the black hole
(M), the accretion efficiency ($\epsilon$) and the luminosity (L) is,

\begin{equation}
L = 1.38\times10^{38} \times \frac {\epsilon M}{M_{\odot}} ~ergs ~ s^{-1}
\end{equation}

where $M$ has units of $M_{\odot}$. Now, if M82 X-1 were an
intermediate mass black hole of, say, a few 1000 M$_{\odot}$, the
value of $\epsilon$ is of order 0.1. The radii at which the disk warps
is then $\sim$ a few 1000 R$_{s}$. On the other hand, if the source
were a stellar-mass black hole, say of 20 M$_{\odot}$, the value of
$\epsilon$ is about 10 which results in warping at radii of a few
R$_{s}$, i.e., the innermost regions of the disk.

The majority of the disk flux is emitted from its innermost regions
($\sim$ a few 10 R$_{s}$), by the gravitational energy loss of the
in-falling material. As noted above there are two possible disk
structures: (1) where the inner disk is warped while the outer disk
remains flat (stellar-mass black hole scenario) or (2) the outer disk
is warped with the inner disk remaining flat (intermediate-mass black
hole scenario). In the first case, as the innermost disk precesses the
X-ray disk flux originating from this region will also modulate at the
precession period thus naturally explaining the observed disk
modulation (Figure 4). In the second case the direct disk emission is
expected to remain constant with the precession period. However, the
disk photons can reflect off the warp in the outer disk and this
reflected component will modulate at the precession period. In this
case the reflection can also produce emission features, viz., Fe
k$\alpha$. The strength of the reflection is proportional to the
projected surface area of the warped disk where reflection
occurs. Therefore any such emission lines would be expected to vary
periodically with the phase of the precession period. The quality of
the current data (Table 1) does not allow us to solve this problem,
however, this should be possible in the near future using
phase-resolved X-ray spectroscopy.

\section{Orbital Scenario and caveats}
While our results show that the 62 day period of M82 X-1 may be due to
a precessing accretion disk they do not yet rule out an orbital
nature. In the standard picture of periodic X-ray modulations from
X-ray binaries, obscuration by, for example, a hot spot at the edge of
the accretion disk (accretion stream interaction site: see Parmar \&
White 1988; Armitage \& Livio 1998) is thought to produce the regular
variations at the orbital period. It is interesting to note that the
phase offset appears to occur just prior to the large flare occurring
around day 1000. If the standard hot spot model is at play here, it is
conceivable that a sudden influx in the accreting material may have
shifted the hot spot and caused an apparent phase shift. Furthermore,
the flux from M82 (Figure 1) is a combination of multiple sources
within {\it RXTE}'s field-of-view (see Figure 1 of Matsumoto et
al. 2001). If the power-law component were dominated by the
contaminating sources one expects it to remain constant with phase.
But, given the fact that there is some evidence for a varying
power-law flux with phase (bottom panel of Figure 4) it is likely that
part of the power-law contribution comes from M82 X-1. This can be
tested with future {\it NuSTAR} observations. While {\it NuSTAR} will
not be able to spatially resolve M82 X-1, energy-dependent surface
brightness modeling similar to that reported by Pasham \& Strohmayer
(2013) can, in principle, constrain its high-energy X-ray spectrum. On
the other hand, the assumption of a thermal accretion disk and a
power-law corona for the X-ray energy spectra of ULXs has been
questioned (e.g., Gladstone et al. 2009).

In summary, our results suggest that the 62 day X-ray period of M82
X-1 may be due to a precessing accretion disk. This hypothesis would
be greatly strengthened if a variation in the observed periodicity can
be confirmed. This can be explored with future monitoring observations
using instruments such as the X-ray telescope on board {\it Swift}.

We thank Craig Markwardt for providing us the
latest {\it RXTE}/PCA background model and the referee
for valuable suggestions that helped us improve the paper.





\begin{thebibliography}

\bibitem[Armitage \& Livio(1998)]{1998ApJ...493..898A} Armitage, P.~J., \& Livio, M.\ 1998, \apj, 493, 898 

\bibitem[Arnaud(1996)]{1996ASPC..101...17A} Arnaud, K.~A.\ 1996, Astronomical Data Analysis Software and Systems V, 101, 17 

\bibitem[Begelman(2002)]{2002ApJ...568L..97B} Begelman, M.~C.\ 2002, \apjl, 568, L97 

\bibitem[Caballero-Garc{\'{\i}}a(2011)]{2011MNRAS.418.1973C} Caballero-Garc{\'{\i}}a, M.~D.\ 2011, \mnras, 418, 1973 

\bibitem[Clarkson et al.(2003)]{2003MNRAS.343.1213C} Clarkson, W.~I., Charles, P.~A., Coe, M.~J., \& Laycock, S.\ 2003, \mnras, 343, 1213 

\bibitem[Colbert \& Mushotzky(1999)]{1999ApJ...519...89C} Colbert, E.~J.~M., \& Mushotzky, R.~F.\ 1999, \apj, 519, 89 

\bibitem[Dewangan et al.(2006)]{2006ApJ...637L..21D} Dewangan, G.~C., Titarchuk, L., \& Griffiths, R.~E.\ 2006, \apjl, 637, L21 

\bibitem[Fabbiano(1989)]{1989ARA&A..27...87F} Fabbiano, G.\ 1989, \araa, 27, 87 

\bibitem[Farrell et al.(2006)]{2006MNRAS.367.1457F} Farrell, S.~A., Sood, R.~K., \& O'Neill, P.~M.\ 2006, \mnras, 367, 1457 

\bibitem[Feng \& Kaaret(2010)]{2010ApJ...712L.169F} Feng, H., \& Kaaret, P.\ 2010, \apjl, 712, L169 

\bibitem[Gladstone et al.(2009)]{2009MNRAS.397.1836G} Gladstone, J.~C., Roberts, T.~P., \& Done, C.\ 2009, \mnras, 397, 1836 

\bibitem[Han et al.(2012)]{2012RAA....12.1597H} Han, X., An, T., Wang, J.-Y., et al.\ 2012, Research in Astronomy and Astrophysics, 12, 1597 

\bibitem[Hopman et al.(2004)]{2004ApJ...604L.101H} Hopman, C., Portegies Zwart, S.~F., \& Alexander, T.\ 2004, \apjl, 604, L101 

\bibitem[Horne \& Baliunas(1986)]{1986ApJ...302..757H} Horne, J.~H., \& Baliunas, S.~L.\ 1986, \apj, 302, 757 

\bibitem[Jain et al.(2010)]{2010MNRAS.409..755J} Jain, C., Paul, B., \& Dutta, A.\ 2010, \mnras, 409, 755 

\bibitem[Kaaret et al.(2006)]{2006ApJ...646..174K} Kaaret, P., Simet, M.~G., \& Lang, C.~C.\ 2006, \apj, 646, 174 

\bibitem[Kaaret \& Feng(2007)]{2007ApJ...669..106K} Kaaret, P., \& Feng, H.\ 2007, \apj, 669, 106 

\bibitem[Kaaret et al.(2009)]{2009ApJ...692..653K} Kaaret, P., Feng, H., \& Gorski, M.\ 2009, \apj, 692, 653 

\bibitem[Katz(1973)]{1973NPhS..246...87K} Katz, J.~I.\ 1973, Nature Physical Science, 246, 87 

\bibitem[King et al.(2001)]{2001ApJ...552L.109K} King, A.~R., Davies, M.~B., Ward, M.~J., Fabbiano, G., \& Elvis, M.\ 2001, \apjl, 552, L109 

\bibitem[K{\"o}rding et al.(2002)]{2002A&A...382L..13K} K{\"o}rding, E., Falcke, H., \& Markoff, S.\ 2002, \aap, 382, L13 

\bibitem[Kotze \& Charles(2012)]{2012MNRAS.420.1575K} Kotze, M.~M., \& Charles, P.~A.\ 2012, \mnras, 420, 1575 

\bibitem[Leahy \& Igna(2010)]{2010ApJ...713..318L} Leahy, D.~A., \& Igna, C.~D.\ 2010, \apj, 713, 318 

\bibitem[Levine et al.(2000)]{2000ApJ...541..194L} Levine, A.~M., Rappaport, S.~A., \& Zojcheski, G.\ 2000, \apj, 541, 194 

\bibitem[Matsumoto et al.(2001)]{2001ApJ...547L..25M} Matsumoto, H., Tsuru, T.~G., Koyama, K., et al.\ 2001, \apjl, 547, L25 

\bibitem[Miller \& Colbert(2004)]{2004IJMPD..13....1M} Miller, M.~C., \& Colbert, E.~J.~M.\ 2004, International Journal of Modern Physics D, 13, 1 

\bibitem[Mucciarelli et al.(2006)]{2006MNRAS.365.1123M} Mucciarelli, P., Casella, P., Belloni, T., Zampieri, L., \& Ranalli, P.\ 2006, \mnras, 365, 1123 

\bibitem[Ogilvie \& Dubus(2001)]{2001MNRAS.320..485O} Ogilvie, G.~I., \& Dubus, G.\ 2001, \mnras, 320, 485 

\bibitem[Parmar \& White(1988)]{1988MmSAI..59..147P} Parmar, A.~N., \& White, N.~E.\ 1988, \memsai, 59, 147 

\bibitem[Pasham \& Strohmayer(2013)]{2013ApJ...764...93P} Pasham, D.~R., \& Strohmayer, T.~E.\ 2013, \apj, 764, 93 

\bibitem[Pasham \& Strohmayer(2013)]{2013ApJ...771...101P} Pasham, D.~R., \& Strohmayer, T.~E.\ 2013, \apj, 771, 101

\bibitem[Portegies Zwart et al.(2005)]{2005Ap&SS.300..247P} Portegies Zwart, S.~F., Dewi, J., \& Maccarone, T.\ 2005, \apss, 300, 247 

\bibitem[Pringle(1996)]{1996MNRAS.281..357P} Pringle, J.~E.\ 1996, \mnras, 281, 357 

\bibitem[Scargle(1982)]{1982ApJ...263..835S} Scargle, J.~D.\ 1982, \apj, 263, 835 

\bibitem[Servillat et al.(2011)]{2011ApJ...743....6S} Servillat, M., Farrell, S.~A., Lin, D., et al.\ 2011, \apj, 743, 6 

\bibitem[Strohmayer \& Mushotzky(2003)]{2003ApJ...586L..61S} Strohmayer, T.~E., \& Mushotzky, R.~F.\ 2003, \apjl, 586, L61 

\bibitem[Strohmayer(2009)]{2009ApJ...706L.210S} Strohmayer, T.~E.\ 2009, \apjl, 706, L210 

\bibitem[Swartz et al.(2011)]{2011ApJ...741...49S} Swartz, D.~A., Soria, R., Tennant, A.~F., \& Yukita, M.\ 2011, \apj, 741, 49 

\bibitem[Verbunt(1993)]{1993ARA&A..31...93V} Verbunt, F.\ 1993, \araa, 31, 93 

\bibitem[Wen et al.(2006)]{2006ApJS..163..372W} Wen, L., Levine, A.~M., Corbet, R.~H.~D., \& Bradt, H.~V.\ 2006, \apjs, 163, 372 

\bibitem[Wolff et al.(2009)]{2009ApJS..183..156W} Wolff, M.~T., Ray, P.~S., Wood, K.~S., \& Hertz, P.~L.\ 2009, \apjs, 183, 156 

\end{thebibliography}
\end{document}